\documentclass[aps,prl,twocolumn,showpacs]{revtex4-1}
\usepackage{epsfig}
\usepackage{bm}
\usepackage{color}
\usepackage{ulem}
\newcommand{\be}{\begin{eqnarray}}
\newcommand{\ee}{\end{eqnarray}}
\newcommand{\ba}{\begin{array}}
\newcommand{\ea}{\end{array}}
\newcommand{\bmat}{\left(\begin{array}}
\newcommand{\emat}{\end{array}\right)}
\newcommand{\no}{\nonumber}

\begin{document}
\title{A useful fundamental speed limit for the imaginary-time Schr\"odinger equation}
\author{Manaka Okuyama$^1$}
\author{Masayuki Ohzeki$^2$}%
\affiliation{%
$^1$Department of Physics, Tokyo Institute of Technology, Oh-okayama, Meguro-ku, Tokyo 152-8551, Japan
}
\affiliation{%
$^2$Graduate School of Information Sciences, Tohoku University, Sendai 980-8579, Japan
}

\date{\today}

\begin{abstract} 
The quantum speed limit (QSL), or the energy-time uncertainty relation, gives a fundamental speed limit for quantum dynamics.
Recently, Kieu [arXiv:1702.00603] derived a new class of QSL which is not only formal but also suitable for actually evaluating the speed limit. 
Inspired by his work, we obtain a similar speed limit for the imaginary-time Schr\"odinger equation.
Using this new bound, we show that the optimal computational time of the Grover problem in imaginary-time quantum annealing is bounded from below by $\log N$, which is consistent with a result of previous study.

\end{abstract}	

\maketitle

\subsection {Introduction}
It is an  essential problem to pursue a limit of physical law.
In the theory of relativity, objects can not exceed the speed of light.
The quantum speed limit (QSL) gives a fundamental limit on time evolution of quantum system and prohibits a state from changing infinitely quickly \cite{MT,ML}.
The existence of limitation in quantum dynamics is very important both theoretically and experimentally, and is still attracting a lot of interest \cite{Lloyd,GLM,Uhlmann,DL,Pfeifer,TEDF,CEPH,DL2,AA,JK,Zwierz,RS,PCCAP,Bhattacharyya,MP,LT,MMGC,Hegerfeldt,CMCFMGS,Luo,DL3,GC}.	

The result of QSL is geometrically clear \cite{AA,PCCAP} and it means that, when the distance between two states is defined properly, the time evolution along the geodesic line is the shortest path.
Although these results are mathematically elegant, it is, in general, very difficult to evaluate the bound for a given time-dependent Hamiltonian and initial state because it requires information on the middle state of time evolution.

On the other hand, there is another class of QSL and  it is not only formal but also suitable for actually evaluating the speed limit \cite{Pfeifer,LGC,Kieu}.
For example, for a given Hamiltonian $\hat{H}(t)$, initial state $|\psi_0\rangle$ and state $|\psi(\tau)\rangle$ following the Schr\"odinger equation, Kieu \cite{Kieu} derived the following inequality
\be
\hbar\| |\psi(\tau)\rangle -e^{ - i\int_0^\tau ds \alpha(s)}  |\psi_0\rangle \| \le \int_0^\tau dt \| (\hat{H}(t)-\alpha(t))|\psi_0 \rangle \|  \label{real-speed-limit} ,
\no\\
\ee
where  $\alpha(t)$ is a time-dependent arbitrary function.
Note that Eq. (\ref{real-speed-limit}) contains only the initial state and given Hamiltonian, and does not require the information on the middle state of the dynamics. This enables us to easily evaluate  the right hand side of Eq. (\ref{real-speed-limit}).
In addition, Eq. (\ref{real-speed-limit}) is not only computable but also tight.
Recent study \cite{Kieu2,Lychkovskiy} shows that, using Eq. (\ref{real-speed-limit}), the optimal computational time of the Grover problem \cite{Grover,RC} is order $\sqrt{N}$ in quantum annealing \cite{KN,FGGS}.

While we have focused on quantum system so far, QSL is not a purely quantum phenomenon and, recently, is extended to classical system such as the classical Liouville equation \cite{SCMC,OO}.
Furthermore, similar speed limits are also obtained for the imaginary-time Schr\"odinger equation such as the classical master equation and the Fokker-Planck equation \cite{OO}.  
In this study, we extended the Kieu bound to the imaginary-time Schr\"odinger equation.	
We obtained a fundamental speed limit for the imaginary-time Schr\"odinger equation which is very similar to the Kieu bound.
However, in the imaginary-time Schr\"odinger equation, the norm of the state is not preserved and it is not clear whether the new bound is tight.
Then, we applied it to the Grover problem in imaginary-time quantum annealing.
Recent study \cite{OOT} shows analytically and numerically that the Grover problem in imaginary-time quantum annealing can be solved by order $\log{N}$.
Here, using our new  bound, we showed that the optimal computational time is  order  $\log{N}$.
This result means that our new bound for the imaginary-time Schr\"odinger equation is also tight and useful.

\subsection{Speed limit for the imaginary-time Schr\"odinger equation}
We consider two imaginary-time Schr\"odinger equations and assume that $\hat{H}$ is a real positive-semidefinite matrix and $|\psi(t)\rangle $ and $|\phi(t)\rangle$ are real vectors,
\be
- \partial_t |\psi(t)\rangle &=&\hat{H}(t)|\psi(t)\rangle \label{Sch1}, 
\\
- \partial_t |\phi(t)\rangle &=&\beta(t) {\bm 1} |\phi(t)\rangle \label{Sch2},
\\
|\psi(0)\rangle&=&|\phi(0)\rangle=|\psi_0 \rangle ,
\ee
where $\beta(t)$ is a time-dependent arbitrary function and $ {\bm 1}$ is the identity matrix.
Taking the difference between Eqs. (\ref{Sch1}) and (\ref{Sch2}), we obtain
\be
\partial_t (|\psi(t)\rangle - |\phi(t)\rangle )&=&-\hat{H}(t)(|\psi(t)\rangle-|\phi(t)\rangle) 
\no\\
&&-(\hat{H}(t)-\beta(t)) |\phi(t)\rangle \label{diff} .
\ee
Considering the distance between $|\psi(t)\rangle $ and $|\phi(t)\rangle$, we obtain
\be
&&\partial_t \| |\psi(t)\rangle - |\phi(t)\rangle \|^2 
\no\\
&=&2 (\langle \psi(t)|-\langle \phi(t)|) \partial_t (|\psi(t)\rangle - |\phi(t)\rangle) \label{diff2} .
\ee
Substituting Eq. (\ref{diff}) into Eq. (\ref{diff2}), we obtain
\be
&&\partial_t \| |\psi(t)\rangle - |\phi(t)\rangle \|^2
\no\\
&=&-2 (\langle \psi(t)|-\langle \phi(t)|) \hat{H}(t)(|\psi(t)\rangle-|\phi(t)\rangle)
\no\\
&&-  2(\langle \psi(t)|-\langle \phi(t)|)(\hat{H}(t)-\beta(t)) |\phi(t)\rangle 
\no\\
&\le&-  2(\langle \psi(t)|-\langle \phi(t)|)(\hat{H}(t)-\beta(t)) |\phi(t)\rangle  
\no\\
&\le&2\| |\psi(t)\rangle-|\phi(t)\rangle\| \| (\hat{H}(t)-\beta(t)) |\phi(t)\rangle  \| \label{diff3} ,
\ee
where we use  $\hat{H}(t)$ being positive-semidefinite  in the first inequality and the second inequality is a result of the Schwarz inequality.
Furthermore, the left hand side of Eq. (\ref{diff3}) can be represented by 
\be
&&\partial_t \| |\psi(t)\rangle - |\phi(t)\rangle \|^2
\no\\
&=&2 \| |\psi(t)\rangle - |\phi(t)\rangle \|  \partial_t \| |\psi(t)\rangle - |\phi(t)\rangle \| .
 \label{diff4} 
\ee
Then, eliminating $ \| |\psi(t)\rangle - |\phi(t)\rangle \|$ from Eqs. (\ref{diff3}) and (\ref{diff4}), we get the following inequality
\be
\partial_t \| |\psi(t)\rangle - |\phi(t)\rangle \| &\le&  \| (\hat{H}(t)-\beta(t)) |\phi(t)\rangle  \| .
\ee
Integrating both the sides with respect to time, we obtain a fundamental speed limit for the imaginary-time Schr\"odinger equation as
\be
\| |\psi(\tau)\rangle -|\phi(\tau)\rangle\| \le \int_0^\tau dt \| (\hat{H}(t)-\beta(t))|\phi(t)\rangle \|  \label{speed-limit} ,
\ee
where $|\phi(t)\rangle=\exp(- \int_0^t ds \beta(s) )|\psi_0\rangle$.
	
Equation (\ref{speed-limit}) is the main result of this Letter which corresponds to Eq.  (\ref{real-speed-limit}).
However, we note that $|\psi(\tau)\rangle$ and $|\phi(\tau)\rangle$ are not normalized, which is a great difference from the case for quantum system. 

\subsection{Speed limit for time-independent system}
First, we consider the time-independent system $\hat{H}(t)=\hat{H}$.
We can evaluate the right hand side of Eq. (\ref{speed-limit}) as 
\be
&&\| (\hat{H}(t)-\beta(t))|\phi(t)\rangle \|
\no\\
&=&\|(\hat{H}-\beta(t))|\psi_0\rangle \| e^{- \int_0^t ds \beta(s) }
\no\\
&=& \sqrt{ \langle \psi_0|\hat{H}^2|\psi_0\rangle- \langle \psi_0|\hat{H}|\psi_0\rangle^2+ (\beta(t) - \langle \psi_0|\hat{H}|\psi_0\rangle)^2 } 
\no\\
&&\times e^{- \int_0^t ds \beta(s) } .
\ee
Setting $\beta(t)= \langle \psi_0|\hat{H}|\psi_0\rangle$, we obtain
\be
&& \int_0^\tau dt\| (\hat{\hat{H}}(t)-\beta(t))|\phi(t)\rangle \|
 \no\\
 &=&\sqrt{ \langle \psi_0|\hat{H}^2|\psi_0\rangle- \langle \psi_0|\hat{H}|\psi_0\rangle^2 } \frac{1-e^{- \tau \langle \psi_0|\hat{H}|\psi_0\rangle }}{\langle \psi_0|\hat{H}|\psi_0\rangle} .
\ee	
Therefore, we find that Eq. (\ref{speed-limit}) is reduced to 
\be
 &&\| |\psi(\tau)\rangle -e^{- \tau \langle \psi_0|\hat{H}|\psi_0\rangle }|\psi(0)\rangle\| 
 \no\\
 &\le& \sqrt{ \langle \psi_0|\hat{H}^2|\psi_0\rangle- \langle \psi_0|\hat{H}|\psi_0\rangle^2 } \frac{1-e^{- \tau \langle \psi_0|\hat{H}|\psi_0\rangle }}{\langle \psi_0|\hat{H}|\psi_0\rangle} .
\ee

\subsection{Speed limit for imaginary-time quantum annealing.}
Next, we consider the following Hamiltonian for application to imaginary-time quantum annealing,
\be
\hat{H}(t)&=&f(t/\tau) \hat{H}_I +g(t/\tau) \hat{H}_P
\\
|\psi_0\rangle&=&|G_I\rangle,
\\
\hat{H}_I |G_I\rangle&=&0,
\ee
where $0\le f(t/\tau), g(t/\tau) \le 1$, $f(0)=g(1)=1, f(1)=g(0)=0$, and $|G_I\rangle$ is the ground state of $\hat{H}_I$.
In quantum annealing, the ground state of the initial Hamiltonian $\hat{H}_I$ is trivial and the ground  state of the target Hamiltonian $\hat{H}_P$ represents an optimal solution of a combinatorial optimization problem. 

We specify the time dependency of $\beta(t)$ as follows,
\be
\beta(t)=\beta_0 g(t/\tau) ,
\ee
where $\beta_0$ is a time-independent any constant.
Then, we find that the right hand side of Eq. (\ref{speed-limit}) is reduced to 
\be
&&\int_0^\tau dt \| (\hat{H}(t)-\beta(t))|\phi(t)\rangle \|
\no\\
&=& \| (\hat{H}_P-\beta_0)|G_I\rangle \|\left( \int_0^\tau dt g(t/\tau) e^{-\beta_0 \int_0^t ds  g(s/\tau) } \right)
\no\\
&=& \sqrt{ \langle G_I|\hat{H}_P^2|G_I\rangle- \langle G_I|\hat{H}_P|G_I\rangle^2 +(\langle G_I|\hat{H}_P|G_I\rangle-\beta_0)^2 } 
\no\\
&&\times  \frac{1-e^{-\beta_0 \int_0^\tau dt  g(t/\tau) }}{\beta_0}  .
\ee 
Therefore, by setting $\beta_0=\langle G_I|\hat{H}_P|G_I\rangle$, we obtain a fundamental speed limit for imaginary-time quantum annealing,
\be
 &&\| |\psi(\tau)\rangle -e^{-\langle G_I|\hat{H}_P|G_I\rangle \int_0^\tau dt  g(t/\tau) }|G_I\rangle\| 
 \no\\
 &\le&\sqrt{ \langle G_I|\hat{H}_P^2|G_I\rangle- \langle G_I|\hat{H}_P|G_I\rangle^2  } \frac{1-e^{-\langle G_I|\hat{H}_P|G_I\rangle \int_0^\tau dt  g(t/\tau) }}{\langle G_I|\hat{H}_P|G_I\rangle}   .
 \no\\ \label{speed-limit-IQA}
 \ee
Although this result is general, it is not clear whether Eq. (\ref{speed-limit-IQA}) is useful for estimating the performance of imaginary-time quantum annealing. 
Then, in the following, we use Eq. (\ref{speed-limit-IQA}) to show that the optimal time of the Grover problem in imaginary-time quantum annealing is order $\log N$.

\subsection{Applicaton to the imaginary-time Grover problem: optimality of $\log N$ }
The Hamiltonian of the Grover problem is given by
\be
 \hat{H}_I  &=& {\bm 1} - |g_I\rangle \langle g_I| ,
\\
\hat{H}_P &=& {\bm 1}-|m\rangle \langle m| .
\ee
In the Grover problem, we start from the ground state of $\hat{H}_I$, which is $|g_I\rangle$, at initial time $t=0$, and hope that the state $|\psi(t)\rangle$ reaches the ground state of  $\hat{H}_P$, which is $|m\rangle$, at final time $t=\tau$.
The relation between $|g_I\rangle$ and $|m\rangle$ is as follows,
\be
\langle g_I | m\rangle&=&\frac{1}{\sqrt{N}} ,
\ee
where $N$ means the size of the problem.
The Hamiltonian $\hat{H}(t)$ is a real positive-semidefinite matrix because the eigenvalues are given by
\be
0 \le E_{\pm}=\frac{1}{2} \left( 1 \pm \sqrt{1-4 f g \left(1- \frac{1}{N} \right) } \right) \le 1 .
\ee
We immediately find that the following relations hold,
\be
\langle g_I|\hat{H}_P|g_I\rangle&=&1-\frac{1}{N} ,
\\
\sqrt{ \langle g_I|\hat{H}_P^2|g_I\rangle- \langle g_I|\hat{H}_P|g_I\rangle^2  }&=&\sqrt{\frac{1}{N}-\frac{1}{N^2}}.
\ee
Setting the initial state $|\phi_0\rangle$ to $|g_I\rangle$, we find that Eq. (\ref{speed-limit-IQA}) is reduced to
\be
&& \| |\psi(\tau)\rangle -e^{-(1-\frac{1}{N}) \int_0^\tau dt  g(t/\tau) }|g_I\rangle\| 
 \no\\
 &\le& \frac{1-e^{-\langle g_I|\hat{H}_P|g_I\rangle \int_0^\tau dt  g(t/\tau) }}{ 1-\frac{1}{N}}  \sqrt{\frac{1}{N}-\frac{1}{N^2}} \label{speed-limit-grover} .
 \ee

We consider the case where the state $|\psi(t)\rangle$ reaches the target state $\||\psi(\tau)\rangle \| \cdot |m\rangle$ at time $\tau$.
In the following, we will find the condition that the computational time $\tau$ must satisfy.
The left hand side of Eq. (\ref{speed-limit-grover}) can be evaluated as
  \be
  &&\| |\psi(\tau)\rangle -e^{-(1-\frac{1}{N}) \int_0^\tau dt  g(t/\tau) }|g_I\rangle\| 
  \no\\
  &=&  \|\| \psi(\tau)\rangle\| \times |m\rangle -e^{-(1-\frac{1}{N}) \int_0^\tau dt  g(t/\tau) } |g_I\rangle \|
\no\\
&=&\left(  \| \psi(\tau)\rangle\|^2 + e^{-2(1-\frac{1}{N}) \int_0^\tau dt  g(t/\tau) }   \right.
\no\\
&&\left. -\frac{2}{\sqrt{N}}\| \psi(\tau)\rangle\|  e^{-(1-\frac{1}{N}) \int_0^\tau dt  g(t/\tau) } \right)^{1/2}
\no\\
&=&\left( \left(1-\frac{1}{N} \right) e^{-2(1-\frac{1}{N}) \int_0^\tau dt  g(t/\tau) }  \right.
\no\\
&&\left. + \left(\frac{1}{\sqrt{N}}e^{-(1-\frac{1}{N}) \int_0^\tau dt  g(t/\tau) }  - \| \psi(\tau)\rangle\| \right)^2 \right)^{1/2}
\no\\
&\ge &\sqrt{ \left(1-\frac{1}{N} \right)e^{-2(1-\frac{1}{N}) \int_0^\tau dt  g(t/\tau) }  }
\no\\
&\ge &\sqrt{ \left(1-\frac{1}{N} \right) e^{-2\tau} } \label{speed-limit-grover2} ,
  \ee
  where we use $0\le   g(t/\tau)\le1$ in the last inequality.
 Then, from Eqs. (\ref{speed-limit-grover} ) and (\ref{speed-limit-grover2}), we find the following inequality
\be
\sqrt{ \left(1-\frac{1}{N} \right) e^{-2\tau} } &\le& \frac{1-e^{-\langle g_I|\hat{H}_P|g_I\rangle \int_0^\tau dt  g(t/\tau) }}{ 1-\frac{1}{N}}  \sqrt{\frac{1}{N}-\frac{1}{N^2}}
\no\\
&\le& \frac{1}{ 1-\frac{1}{N}}  \sqrt{\frac{1}{N}-\frac{1}{N^2}} ,
 \ee
  where we use also the fact that $0\le  g(t/\tau)\le1$ in the last inequality.
Therefore, taking a large limit of $N$, we find
 \be
 \tau \ge  \frac{1}{2}\log (N) ,
 \ee
which is the necessary condition satisfied by the computational time $\tau$.
On the other hand, Ref. \cite{OOT} finds the explicit schedule which can solve the Grover problem by order $\log{N}$ in imaginary-time quantum annealing.
Combining these results, we conclude that the optimal computational time of the Grover problem is order $\log{N}$ in imaginary-time quantum annealing.

\subsection{ Conclusions}
We have provided a fundamental speed limit for the imaginary-time Schr\"odinger equation.
We apply it to imaginary-time quantum annealing, and show that the optimal time of the imaginary-time Grover problem is order $\log N$, which is consistent with the previous study \cite{OOT}.
This result means that the new bound (\ref{speed-limit}) is not only computable but also tight.

In real-time dynamics, the schedule obtained from the adiabatic theorem is optimal \cite{RC} and it is possible to prove the optimality of order $\sqrt{N}$ by the fundamental speed limit (\ref{real-speed-limit}).
On the other hand, in imaginary-time dynamics, the adiabatic theorem is merely a  sufficient condition, and the transition from  excited states to the ground state strongly influences and can not be ignored.
Thus, the imaginary time adiabatic theorem \cite{MN,KN2} does not give the optimal schedule \cite{OOT}.
Even in such a case, the fundamental speed limit (\ref{speed-limit}) can prove the optimality of order $\log N$.
This result means that the adiabatic time evolution has nothing to do with the optimality in imaginary-time dynamics, although the adiabatic time evolution is closely related to the optimality in real-time dynamics.
In addition, Ref. \cite{OOT} pointed out that the imaginary-time annealing is not physically realistic.
Our result shows that there is a fundamental limit even in such non-physical systems.

Although we have focused on imaginary-time quantum annealing which corresponds to population annealing \cite{HI}, it is also expected that the new bound can be applied to estimate the performance of the classical master equation and the Fokker-Planck equation.
It is a future problem to apply the new bound to other classical algorithms.

The authors are deeply grateful to Shuntaro Okada for useful discussions.
M. Okuyama was supported by JSPS KAKENHI Grant No. 17J10198.
M. Ohzeki was supported by ImPACT Program of Council for Science, Technology and Innovation (Cabinet Office, Government of Japan) and JSPS KAKENHI No. 16K13849, No. 16H04382, and the Inamori Foundation.


\end{document}